\documentclass[conference]{IEEEtran}

\ifCLASSINFOpdf 
\else
\fi

\usepackage{amssymb}
\usepackage{amsmath}
\usepackage{bm}
\usepackage{graphicx}
\usepackage{xcolor}
\usepackage{import}
\usepackage[utf8]{inputenc}
\usepackage{listings}
\usepackage[noend]{algpseudocode}
\usepackage{url}
\usepackage{multirow}
\usepackage[font={small,it}]{caption}
\usepackage[bookmarks=false]{hyperref} 
\usepackage{array}
\usepackage{rotating}
\usepackage{makecell}
\usepackage{colortbl}
\usepackage{enumerate}
\usepackage{booktabs}
\usepackage{subfig}
\usepackage{paralist}
\usepackage[ruled,vlined]{algorithm2e}
\usepackage[export]{adjustbox}

\usepackage[bottom]{footmisc} 
\usepackage{ulem}

\usepackage{mathptmx}
\newcommand{\subparagraph}{} 

\algrenewcommand\algorithmiccomment[1]{\hfill \textcolor{gray}{$\triangleright$ \textit{#1}}}

\DeclareUnicodeCharacter{00A0}{ }

\makeatletter
\g@addto@macro\normalsize{
  \setlength\abovedisplayskip{4pt}
  \setlength\belowdisplayskip{4pt}
  \setlength\abovedisplayshortskip{4pt}
  \setlength\belowdisplayshortskip{4pt}
}
\makeatother

\renewcommand{\tablename}{Tab.}

\renewcommand{\figurename}{Fig.}

\renewcommand{\paragraph}[1]{\vspace{0.1cm}\noindent{\bf #1.}}

\usepackage{makeidx}
\usepackage[
    nogroupskip, 
    nonumberlist, 
    ]{glossaries-extra}

\makeindex{}
\makeglossaries{}

\setabbreviationstyle[acronym]{long-short}

\newacronym[plural=DAGs]{dag}{DAG}{Directed Acyclic Graph}
\newacronym{iot}{IoT}{Internet of Things}
\newacronym{pow}{PoW}{Proof of Work}
\newacronym{btc}{BTC}{Bitcoin}

\newglossaryentry{block-creation-rate}
{
    name={\ensuremath{\lambda}},
    description={
            Block creation rate
        }
}
\newglossaryentry{honest-miner}
{
    name={\ensuremath{\mathbb{H}}},
    description={
            Honest miner
        }
}
\newglossaryentry{malicious-miner}
{
    name={\ensuremath{\mathbb{M}}},
    description={
            Malicious miner
        }
}
\newglossaryentry{adversarial-mining-power}
{
    name={\ensuremath{\kappa}},
    description={
            Adversarial mining power
        }
}
\newglossaryentry{network-propagation-delay}
{
    name={\ensuremath{\tau}},
    description={
            Network propagation delay of blocks
        }
}
\newglossaryentry{gap-parameter}
{
    name={\ensuremath{c}},
    description={
            Gap parameter in PHANTOM
        }
}
\newglossaryentry{discount-function}
{
    name={\ensuremath{\gamma}},
    description={
            Discount function in PHANTOM
        }
}
\newglossaryentry{phantom-block}
{
    name={\ensuremath{A}},
    description={
            A block in the PHANTOM solution
        }
}
\newglossaryentry{time}
{
    name={\ensuremath{t}},
    description={
            Time
        }
}
\newglossaryentry{eulers-number}
{
    name={\ensuremath{e}},
    description={
            Euler's number
        }
}

\glssetcategoryattribute{general}{dualindex}{true}

\IEEEoverridecommandlockouts

\begin{document}
\title{DAG-Oriented Protocols PHANTOM and GHOSTDAG under Incentive Attack via Transaction Selection Strategy
\thanks{$\copyright$ 2021 IEEE.  Personal use of this material is permitted.  Permission from IEEE must be obtained for all other uses, in any current or future media, including reprinting/republishing this material for advertising or promotional purposes, creating new collective works, for resale or redistribution to servers or lists, or reuse of any copyrighted component of this work in other works.}}

\author{
    \IEEEauthorblockN{
        Martin Perešíni\IEEEauthorrefmark{1},
        Federico Matteo Benčić\IEEEauthorrefmark{2},
        Kamil Malinka\IEEEauthorrefmark{1} and
        Ivan Homoliak\IEEEauthorrefmark{1}
    }
    \IEEEauthorblockN{
        \href{mailto:iperesini@fit.vutbr.cz}{iperesini@fit.vutbr.cz},
        \href{mailto:federico-matteo.bencic@fer.hr}{federico-matteo.bencic@fer.hr},
        \href{mailto:malinka@fit.vut.cz}{malinka@fit.vut.cz},
        \href{mailto:ihomoliak@fit.vutbr.cz}{ihomoliak@fit.vutbr.cz}
    }
    \IEEEauthorblockA{
        \IEEEauthorrefmark{1}Brno University of Technology, Faculty of Information Technology
    }
    \IEEEauthorblockA{
        \IEEEauthorrefmark{2}University of Zagreb, Faculty of Electrical Engineering and Computing
    }
}

\maketitle

\begin{abstract}
	In response to the bottleneck of processing throughput inherent to single chain PoW blockchains, several proposals have substituted a single chain for Directed Acyclic Graphs (DAGs). In this work, we investigate two notable DAG-oriented designs. We focus on PHANTOM (and its optimization GHOSTDAG), which proposes a custom transaction selection strategy that enables to increase the throughput of the network. 
	However, the related work lacks a thorough investigation of corner cases that deviate from the protocol in terms of transaction selection strategy. 
	Therefore, we build a custom simulator that extends open source simulation tools to support multiple chains and enables us to investigate such corner cases.
	Our experiments show that malicious actors who diverge from the proposed transaction selection strategy make more profit as compared to honest miners. 
	Moreover, they have a detrimental effect on the processing throughput of the PHANTOM (and GHOSTDAG) due to same transactions being included in more than one block of different chains.
	Finally, we show that multiple miners not following the transaction selection strategy are incentivized to create a shared mining pool instead of mining independently, which has a negative impact on decentralization.
\end{abstract}

\section{Introduction}\label{sec:intro}
Blockchains have become popular due to several interesting properties they offer, such as decentralization, immutability, availability, and transparency. Thanks to these properties, blockchains have been adopted in various fields, such as finance, supply chains, identity management, \gls{iot}, record management systems, etc.

Nonetheless, blockchains inherently suffer from the processing throughput bottleneck, as consensus must be reached for each block within the chain. One approach to solving this problem is to increase the rate at which blocks are created. However, this approach has its drawbacks. If blocks are not propagated through the network before a new block (or blocks) is created, a \textit{soft fork} can occur, where two blocks reference the same parent block. A soft fork resolves itself, and thus only one block is eventually accepted as valid. All other blocks and transactions they contain are discarded (i.e., \textit{orphaned}). As a result, consensus nodes that created orphaned blocks waste their resources and do not get rewarded for their efforts. This is problematic because it may discourage consensus nodes from participating in the consensus protocol.

\begin{figure}[t]
	\centering
	\includegraphics[width=0.49\textwidth]{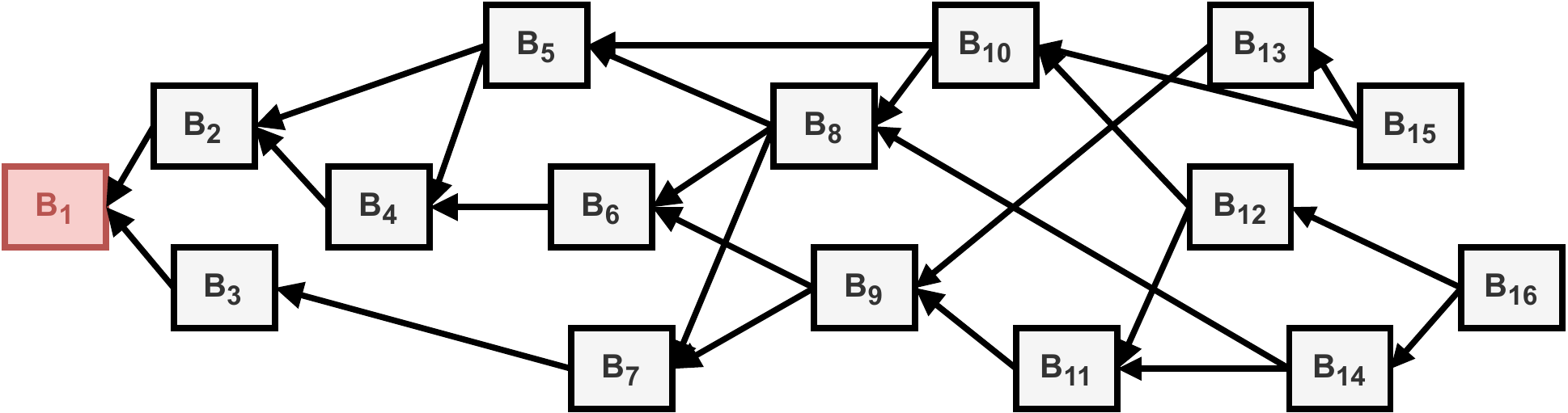}
	\caption{A DAG-oriented blockchain.}\label{fig:dag-chain}
\end{figure}

As a response to the above issue, several proposals~\cite{sompolinsky2013GHOST,sompolinsky2016spectre} have substituted a single chaining data structure for \glspl{dag}, as displayed in \autoref{fig:dag-chain}. 
Such a structure can maintain multiple interconnected chains and thus increase processing throughput. The assumption of some \gls{dag}-oriented solutions is to abandon transaction selection based on the highest fee, since this approach increases the probability that the same transaction is included in more than one block (hereafter \textit{transaction collision}). Instead, these solutions use a random transaction selection strategy to avoid transaction collisions.

However, no one has yet analyzed the performance and robustness of \gls{dag}-oriented approaches within an empirical study assuming real-world conditions and adversarial setting. In this work, we focus on the impact of malicious actors in blockchain networks deploying \gls{dag}-oriented designs. In particular, we study the situation where an attacker (or attackers) deviate from the protocol by not following the random transaction selection strategy that is assumed by a few \gls{dag}-oriented approaches (i.e.,~\cite{sompolinsky2013GHOST,sompolinsky2016spectre}).
We conjecture that it might have two major consequences. First, such an attacker can earn greater rewards compared to honest participants. Second, such an attacker has a detrimental effect on transaction throughput, as \textit{transaction collision} is increased. We verify our hypothesis using an open-source simulation tool Bitcoin mining simulator~\cite{gavinsimulator}, modified to our needs.
We focus on a case study of a selected \gls{dag}-oriented solution called PHANTOM~\cite{sompolinsky2020phantom}.

\paragraph{Contributions}
The contributions of this work are as follows:

\begin{enumerate}
	\item We analyze DAG-oriented protocols, namely PHANTOM~\cite{sompolinsky2020phantom}, GHOSTDAG~\cite{sompolinsky2020phantom}, and select them as the subject of our research due to their practical features.
	\item We show that a malicious actor who selects transactions based on the highest fee in PHANTOM and GHOSTDAG has a significant advantage in making profits compared to honest participants who select transactions randomly.
	\item We show that multiple such malicious actors significantly decrease transaction throughput by increasing the transaction collision rate.
	\item We show that if the block creation rate is increased while there are malicious actors present in the network, the transaction collision rate increases significantly.
	\item We show that malicious actors have a significant incentive to join a mining pool, which as a consequence degrades the decentralization property of the blockchain with PHANTOM or GHOSTDAG protocol employed.
\end{enumerate}

\section{Background}\label{sec:background}

We establish preliminary terms and definitions that will be used throughout this work.

\paragraph{Blockchain}
The blockchain is a tamper-resistant data structure in which data records (i.e., blocks) are linked using a cryptographic hash function, and each new block must be agreed upon by participants (a.k.a., miners or consensus nodes) running a consensus protocol.

\begin{figure}[t]
	\centering
	\includegraphics[width=0.49\textwidth]{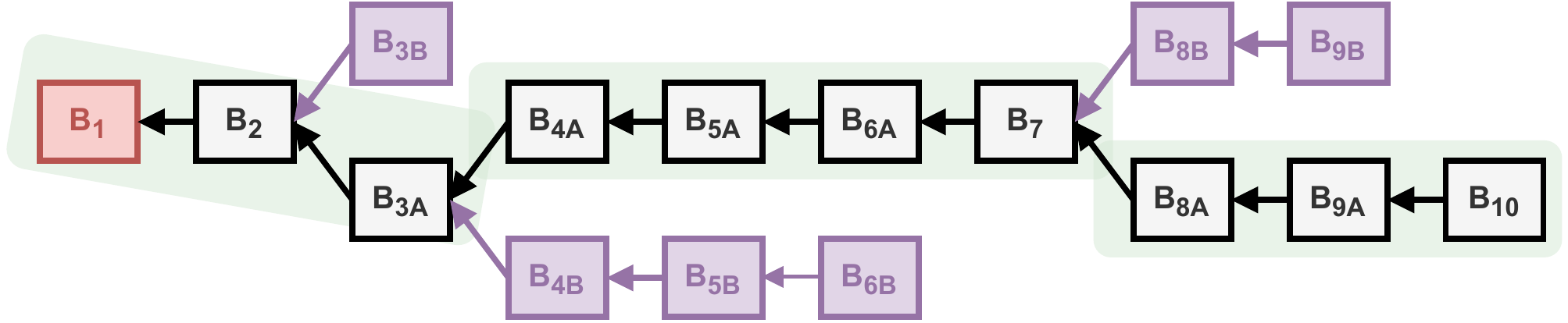}
	\caption{The longest-chain rule with orphaned blocks (purple) graphically displayed in Nakamoto consensus.}\label{fig:longest-chain}
\end{figure}

\begin{figure}[t]
	\centering
	\includegraphics[width=0.47\textwidth]{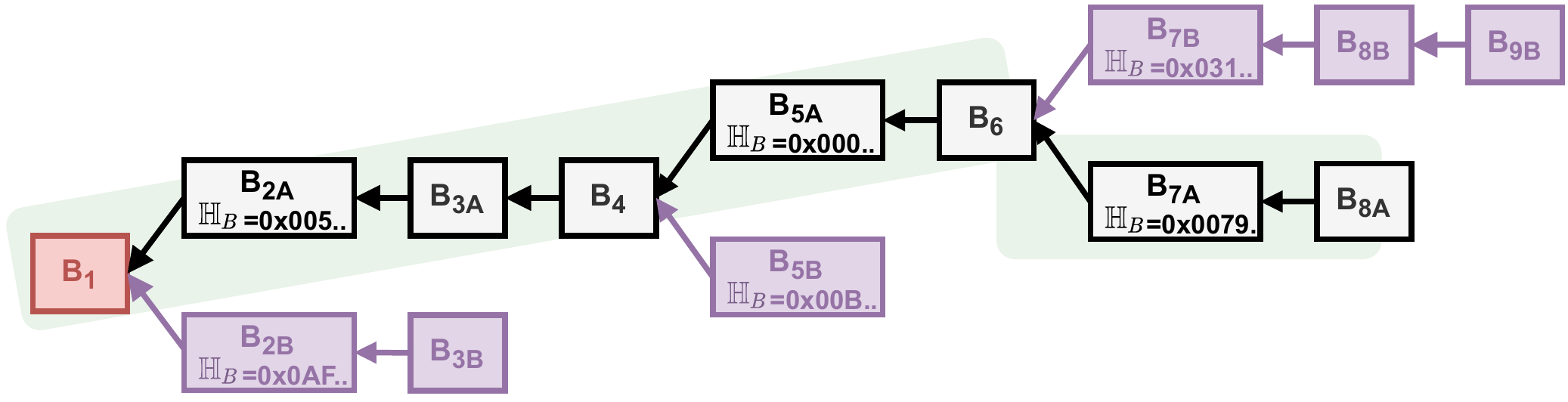}
	\caption{The strongest-chain rule with the main chain depicted in green and orphaned blocks in purple. The hash of the current block is denoted as \(\mathbb{H}_{B}\).}\label{fig:strongest-chain} 
\end{figure}

\paragraph{Nakamoto Consensus}
In the Nakamoto consensus~\cite{nakamoto2008bitcoin}, blocks contain the finite set of transactions, and each block contains the hash of the previous block, which ensures the immutability of the history. 
Nakamoto consensus uses a single chain to link the blocks, while \gls{pow} algorithm is used to establish consensus, which is a mathematical puzzle of cryptographic zero-knowledge hash proof, where one party proves to others that it has spent a certain computational effort, and thus is entitled to be a leader of the round who produces a block. 
This effort represents finding a value below a certain threshold (i.e., determined by the \textit{difficulty} parameter), which is computationally intensive. On the other hand, checking the correctness of the puzzle requires negligible computational effort. The order of blocks\footnote{And consequently transactions.} in \gls{btc} was originally determined using the longest chain rule (see \autoref{fig:longest-chain}). However, this rule was later replaced in favour of the strongest chain rule (see \autoref{fig:strongest-chain}), which takes into account the accumulated difficulty of the \gls{pow} puzzle.

\paragraph{Fees and Rewards}
Miners who create new blocks are rewarded with block rewards. Block rewards refer to new crypto tokens (e.g., \gls{btc}) awarded by the blockchain network. It is assumed that miners earn profits proportionally to their consensus power (e.g., mining power). Another source of crypto-token income for active miners are \textit{transaction fees}, which are awarded to the miner who includes the corresponding transaction in a newly mined block. Transaction fees are paid by clients who deliberately choose the amount based on the priority of a transaction. To maximize profit, miners use a transaction selection mechanism that prioritizes the transactions with the highest fees for inclusion in a block. 

\paragraph{Mempool}
The mempool is a data structure containing transactions that can potentially be included (i.e., mined) in a block by any consensus node. A new transaction is `broadcast', i.e., sent from a client to its peers, which in turn forward the transaction to their peers. This process is repeated until the transaction has propagated throughout the network. Since transactions are not immediately published (processed) on the blockchain, the mempool may vary slightly from node to node until all transactions have completed their propagation.

\paragraph{Block creation time}
In \gls{btc}, there is a default block creation time \gls{block-creation-rate} set to create a new block every \(10\) minutes (i.e., every \(600\) seconds). This parameter is derived directly from the network difficulty, which changes over time, and it is adjusted every \(2016\) blocks to fit target value of 10 minutes (i.e., approximately every two weeks). According to Gervais et al.~\cite{gervais2016security}, the current stale block rate of \gls{btc} is \(0.41\%\). Other sources~\cite{Decker_13-thieee, GOBEL201623} state the values around \(0.5-1\%\), which is considered negligible. We assume that the mathematical model corresponding to the block creation time of \gls{btc} is an exponentially distributed random variable, where the time between two consecutive blocks is expressed as 
\begin{equation}
	f_{\mathbb{T}}(\gls{time})=\Lambda \gls{eulers-number}^{-\Lambda \gls{time}}, 
\end{equation}
where \(\Lambda = \frac{1}{\gls{block-creation-rate}}\)~\cite{bowden2018block, grunspan2020mathematics}, \gls{eulers-number} is the Euler's number and \gls{time} is time in seconds.

\section{DAG-Oriented Solutions}\label{sec:dag-oriented-solutions}

Although there are some \gls{dag}-oriented designs that do not address the problem of increasing transaction throughput (e.g., IoTA~\cite{silvano2020iota}, Nano~\cite{Lemahieu2018NanoA}, Byteball~\cite{Churyumov2016}), 
we focus on the solutions addressing this problem, such as GHOSTDAG, PHANTOM~\cite{sompolinsky2013GHOST}, and SPECTRE~\cite{sompolinsky2016spectre}. 

Since SPECTRE is a theoretical design whose complexity of recursively establishing consensus on the blocks it contains is impractical, we focus and describe PHANTOM in the scope of this paper. Furthermore, the use of PHANTOM is considered impractical for efficient use~\cite{sompolinsky2020phantom}, because it requires the solution of a NP-hard problem\footnote{Maximum \textit{k}-cluster SubDAG problem}, the authors of PHANTOM have developed a greedy algorithm called GHOSTDAG, which is more suitable for implementation. We emphasize that our research results are applicable to GHOSTDAG as well as PHANTOM. In the context of this work, the Maximum \textit{k}-cluster SubDAG problem and block ordering are not crucial, so they have been abstracted in our simulations for simplicity.

\subsection{PHANTOM}
The PHANTOM protocol~\cite{sompolinsky2020phantom} is a generalization of Nakamoto's longest chain protocol. While in Nakamoto consensus, each block contains a hash of one previous block in the chain it extends, PHANTOM organizes blocks in the manner of a \gls{dag}. As a result, each block may contain multiple hash references to predecessors. The key idea of PHANTOM is that it performs a complete ordering of all blocks (and thus transactions). Unlike the Nakamoto consensus, which discards blocks that are not on the main chain (i.e., orphan blocks), PHANTOM includes these blocks in a \gls{dag} structure, 
except for attacker-created blocks, which would be weakly connected to the \gls{dag} ledger. 

One notable difference from the Nakamoto consensus is that PHANTOM does not explicitly define a scheme for block rewards and only considers transaction fees. As mentioned in \autoref{sec:background}, miners in blockchain networks such as \gls{btc} use a well-known transaction selection mechanism that maximizes profit by selecting transactions with the highest fees. In the context of this paper, we refer to this strategy as the \textit{rational strategy}. Instead of the rational transaction selection strategy, PHANTOM uses a \textit{randomized strategy} proposed by the authors of inclusive blockchain protocols~\cite{lewenberg2015inclusive}. In the randomized strategy, the miners do not take into account the fees of the transactions and instead select transactions randomly. In this way, the authors aim to eliminate transaction collision within parallel blocks of a \gls{dag} structure, thus using the parallel blocks as a means to increase the transaction throughput of the protocol.

The incentive scheme of the protocol revolves around rewarding all miners who include a new transaction within a new block \gls{phantom-block}, while assuming that transactions in the parallel blocks are unique~\cite{lewenberg2015inclusive} and due to a DAG-based structure will not be discarded as in single-chain blockchains.\footnote{I.e., duplicate transactions are not rewarded.} 
In this way, poorly connected miners who were not fast enough to publish a new block would still receive rewards. 
However, such an incentive scheme must be constructed with care, as sidechain blocks can also be the result of a malicious attack.

Therefore, the reward a miner receives for publishing \gls{phantom-block} is indirectly proportional to the delay at which \gls{phantom-block} was referenced by the main chain. For this reason, the protocol defines a measure of the delay in publishing \gls{phantom-block} with respect to the main chain, called the \textit{gap parameter} \gls{gap-parameter}. The amount by which the reward is reduced is determined by the discount function \gls{discount-function}, where \(\gls{discount-function}(\gls{gap-parameter}(\gls{phantom-block})) \in [0,1]\) and \gls{discount-function} is weakly decreasing.\footnote{I.e., later inclusion of the side-chain block imposes lower reward.} Finally, the miner is rewarded for the inclusion of transactions in \gls{phantom-block} using the \textit{Payoff Function}. Specifically, the miner gets rewarded for all transactions that were first contained in \gls{phantom-block} and after the discount function \gls{discount-function} was applied to the respective transaction fees.

\paragraph{Comparison with Nakamoto Consensus}
As compared to Nakamoto consensus, PHANTOM brings the following changes:
\begin{compactitem}
	\item \gls{dag}-oriented design instead of a single chain (enabling parallel blocks),
	\item randomized transaction selection strategy.
	\item a reduction of the rewards for transactions with regards to the delay in publishing \gls{phantom-block} with respect to the main chain.
\end{compactitem}

\subsection{GHOSTDAG}
The use of PHANTOM is considered impractical for efficient use~\cite{sompolinsky2020phantom}, because it requires the solution of a NP-hard problem (Maximum \textit{k}-cluster SubDAG problem). In response, the authors of PHANTOM have developed a greedy (heuristic) algorithm called GHOSTDAG, which is more suitable for implementation. 

\section{Problem Definition}\label{sec:problem}
Let there be a PoW blockchain network that uses the Nakamoto consensus and consists of honest and malicious miners, with the malicious miners holding a minority of the mining power. We refer to honest miners as \gls{honest-miner}, malicious miners as \gls{malicious-miner}, and their adversarial mining power as \gls{adversarial-mining-power} (i.e., a fraction of the total mining power). Then, we denote the network propagation delay in seconds as \gls{network-propagation-delay}, and the block creation time in seconds as \gls{block-creation-rate}. 
We assume that the minimum value of \gls{block-creation-rate} is constrained by the network propagation delay of the blockchain network, denoted as \gls{network-propagation-delay}.

\begin{figure}[t]
	\centering
	\includegraphics[width=\linewidth]{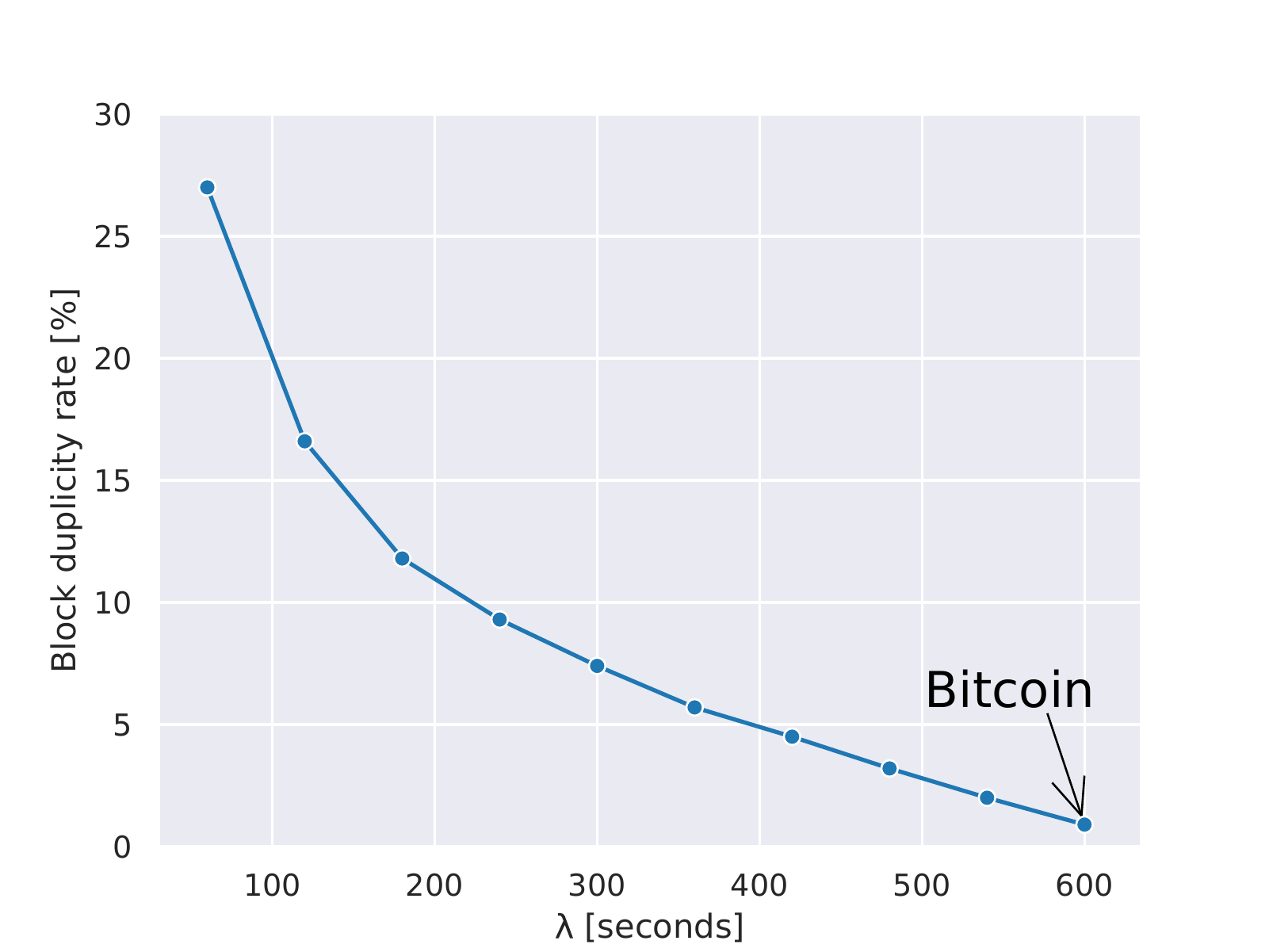}
	\caption{Relationship between orphan block rate (i.e., percentage of duplicate blocks) and block creation time \gls{block-creation-rate}.}\label{fig:lambda-graph}
\end{figure}

It is well-known Nakamoto-style blockchains generate stale blocks (a.k.a., orphan blocks). As a result, a fraction of the mining power is wasted. The rate at which stale blocks are generated increases when the block creation time \gls{block-creation-rate} is decreased (see \autoref{fig:lambda-graph}), which is one of the reasons why Bitcoin maintains a relatively high block creation time of 600 seconds.

Furthermore, let the blockchain utilize the \gls{dag}-oriented solution PHANTOM~\cite{sompolinsky:cryptoeprint:2018:104} than enables to significantly reduce \gls{block-creation-rate} and thus increase transaction throughput since blocks are produced more often.

\paragraph{Assumptions for Our Attacks}
Alike PHANTOM, we assume that incentive scheme does not utilize block rewards but transaction fees.
Then, let us assume that no stale blocks are the result of a malicious action, and the malicious miners may only choose a different transaction selection strategy to make more profit than honest miners. This assumption has two major consequences:
\begin{compactitem}
	\item 
	First, the protocol can safely use the naive approach where the miner of the block \gls{phantom-block} gets rewarded for all unique transactions in \gls{phantom-block}, regardless of the distance of \gls{phantom-block} from the main chain. In other words, for each block \gls{phantom-block}, the discount function does not penalize a block according to its gap parameter \(\gls{gap-parameter}(\gls{phantom-block})\), i.e. \(\gls{discount-function}(\gls{gap-parameter}(\gls{phantom-block})) = 1\). 
	As a consequence, such a setting maximizes the profit from transaction fees that honest miners are expected to make when using a random transaction selection strategy.

	\item 
	Second, the existence of malicious miners using other than random transaction selection strategy (e.g., rational strategy) increases the transaction collision rate. Consequently, there are fewer unique transactions in each block, and the transaction throughput is reduced.		
\end{compactitem}

\paragraph{Identified Problems}
Neither the authors of PHANTOM and GHOSTDAG nor related work studied the impact of malicious miners deviating from the random transaction selection strategy and the effect they might have on the throughput of the protocol as well as a fair distribution of earned rewards (in the form of transaction fees). Therefore, we focus on studying the transaction selection strategy. Since honest nodes in the PHANTOM protocol select transactions from the mempool using the random selection strategy, one might wonder what happens in the case of the attacker who selects transactions based on the maximum fees (i.e., the rational strategy). 
Our hypothesis is that a rational transaction selection strategy will decrease the relative profit of honest miners as well as transaction throughput.

\section{Setup \& Simulation Model}\label{sec:verification} 
We created a simulation model that enables the execution of various experiments aimed at investigation of the PHANTOM protocol's behavior under attacks related to the potential problems identified in \autoref{sec:problem}.

\subsection{Network Topology}
Our intention was to create a network topology that is convenient for the simulation and encompasses some aspects of the real-world blockchain network. 
In particular, we were interested in emulating the network propagation delay to be similar as in Bitcoin. 
We assumed that Bitcoin contains around 8500 consensus nodes~\cite{Park2019NodesIT}, while each node usually has eight peers (i.e., a default value in Bitcoin Core). 
Therefore, the maximum number of hops that a gossiped message requires to reach all consensus nodes in the network is $\sim4.35$ (i.e., $log_8(8500)$).
Moreover, if we were to assume that there are two to three times more independent blockchain clients who are not consensus nodes, then this number would be increased by 1.

\begin{figure}[t]
	\centering
	\includegraphics[width=.66\linewidth]{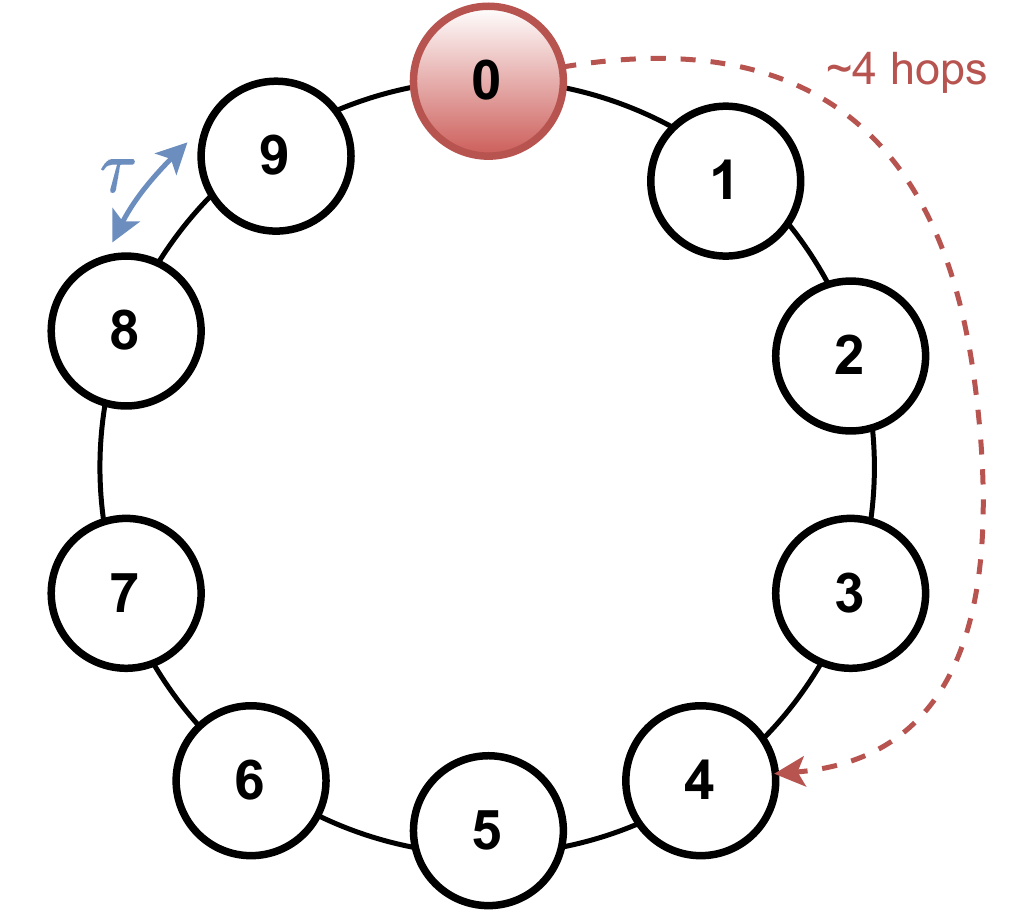}
	\caption{The network topology utilized in our simulations.}
	\label{network-topology}
\end{figure}

Therefore, we utilized the ring network topology with 10 consensus nodes (see \autoref{network-topology}), which sets the maximum value of hops required to propagate a message to 5. 
In this way, we could adjust the adversarial mining power with the granularity of 10\%.

\subsection{Simulator}
There are many simulators~\cite{9366733} aiming at blockchain protocols, mainly focusing on network delays, different consensus protocols, behavior of specific attacks (e.g., SimBlock~\cite{Aoki2019SimBlockAB}, Blocksim~\cite{BlockSim:Alharby}, Bitcoin-Simulator~\cite{Bitcoin-Simulator} etc.).
However, none of these simulators was sufficient for our purposes due to missing support for multiple chains and incentive schemes specific to assumed DAG protocols.
To verify our hypothesis, we have built a simulator that focuses on the mentioned problems of DAG protocols.
In detail, we started with the Bitcoin mining simulator~\cite{gavinsimulator}, which is a discrete event simulator for the PoW mining process in a single chain, and it enables simulation of network propagation delay within a specified network topology.
We extended the Bitcoin mining simulator to support the simulation of (abstracted) multiple chains of DAG-based blockchains while monitoring transaction duplicity, throughput, and relative earned profit with regard to the used mining power.
The simulator is written purely in \verb!C++!.
The implementation utilizes the Boost library~\cite{boost} for better performance and special structures used for simulation, especially the multi-index mempool~\cite{boost2} structure, which enables effective management of the mempool in the case of any transaction selection strategy.\footnote{Rational transaction strategy requires a mempool with transactions ordered by fees, while random transaction strategy requires the hash-map data structure. Therefore, they are hard to utilize at the same time.}

In addition, we have improved and added more simulation complexity to simulate each block, including the corresponding transactions (as opposed to simulating only the number of transactions in a block~\cite{gavinsimulator}), added a custom process that creates transactions and broadcasts them to the network of miners, and most importantly, we have implemented two different transaction selection strategies -- a rational strategy and random strategy.
All important parameters of our simulations and their ranges are summarized in \autoref{tab:params-simulation}.

\begin{table}[b]
	\vspace*{1em}
	\centering
	\begin{tabular}{r l }
		\toprule
		Parameter & Value \\
		\midrule
		block creation rate ($\lambda$) &  10 to 600 \\ 
		network propagation delay ($\tau$)  & 5 seconds \\ 
		$\#$ of total blocks & 10000 \\ 
		$\#$ of miners & 10 \\ 
		network topology & ring \\
		$\#$ of transactions in block & 100 \\
		maximum size of mempool & 10000 transactions \\
		distribution of transaction fees & exponential ($\lambda$ = 150) \\
		mining strategies & rational or random \\
		mining power of each miner & 0 to 49\% \\
		\bottomrule
	\end{tabular}
	\caption{The summary of parameters in our simulations.}\label{tab:params-simulation}
\end{table}

\section{Evaluation}\label{sec:evaluation} 
During the evaluation, we executed three experiments that investigated the relative profit of malicious miners, transaction collision rate, and throughput under different settings.
In all experiments, honest miners used the random transaction selection strategy (as proposed in PHANTOM and GHOSTDAG), while malicious miners used the rational transaction selection strategy. Unless stated otherwise, each miner held \(10\%\) of the mining power.

\begin{figure}[t]
	\centering
	\includegraphics[width=\linewidth]{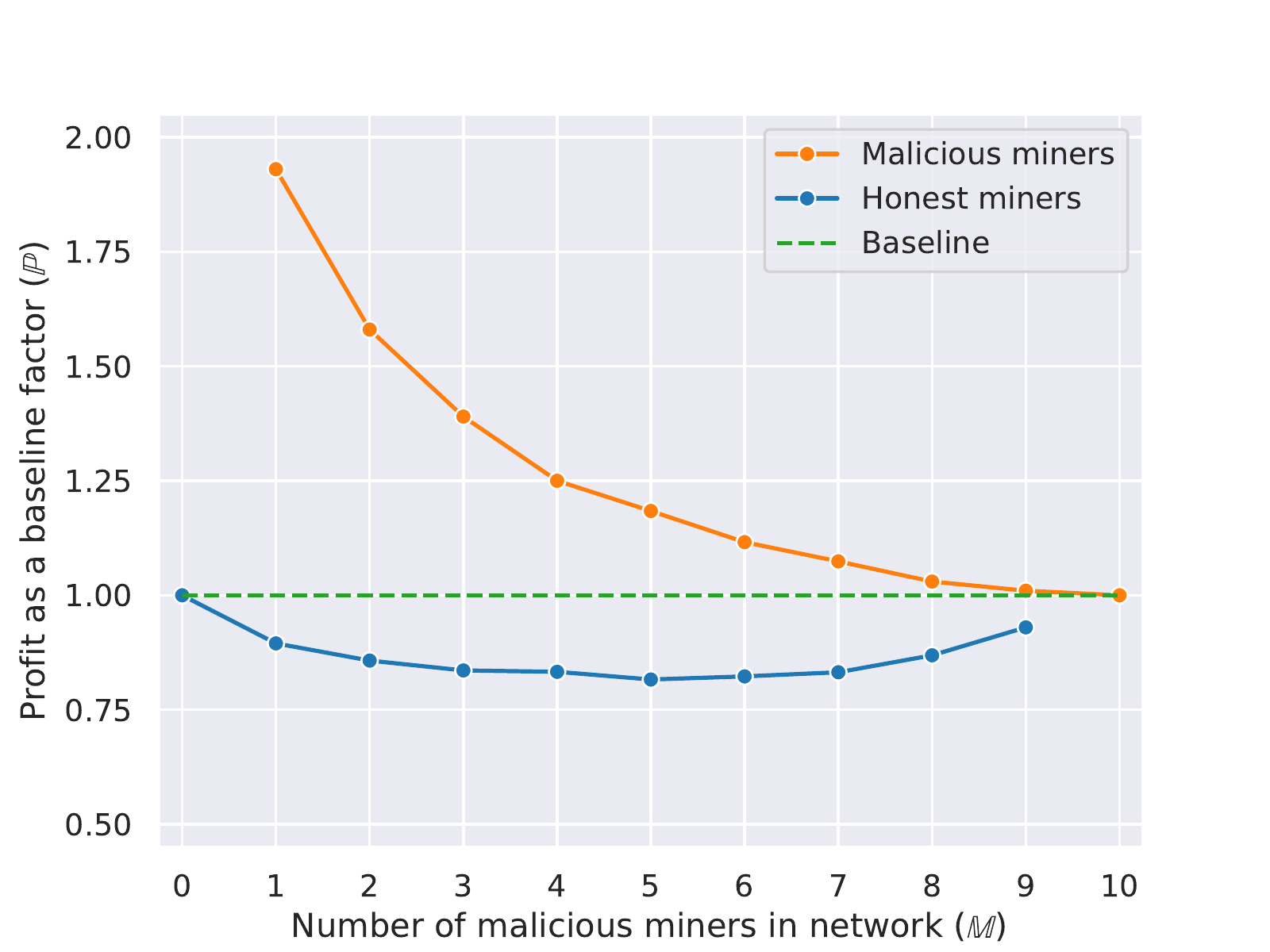}
	\caption{The average profit per honest and malicious miner based on the number of malicious miners (equipped with \(10\%\) of mining power). The baseline shows the expected average profit of an honest miner with \(10\%\) of the mining power.}\label{fig:malicious-miners-earn-more-profit} 
\end{figure}

\subsection{Experiment I} 
\paragraph{Goal} The goal of this experiment is to investigate relative earned profits of malicious miners following rational transaction selection strategy in contrast to honest miners who follow random strategy.

\paragraph{Methodology and Results}
The block creation time was kept constant within this experiment (\(\gls{block-creation-rate}=20\)), and we experimented with the number of malicious miners\footnote{Each holding \(10\%\) of the total mining power.} in the network and monitored the transaction collision rate for each setting.\footnote{Expressed as a percentage of all transactions mined vs.\ the capacity of all blocks during a single simulation run.} 
The results are depicted in \autoref{fig:malicious-miners-earn-more-profit}, 
and they show that with a single malicious miner, the average profit of honest miners is roughly halved as compared to the average profit of the malicious miners. 

\begin{figure}[b]
	\centering
	\includegraphics[width=\linewidth]{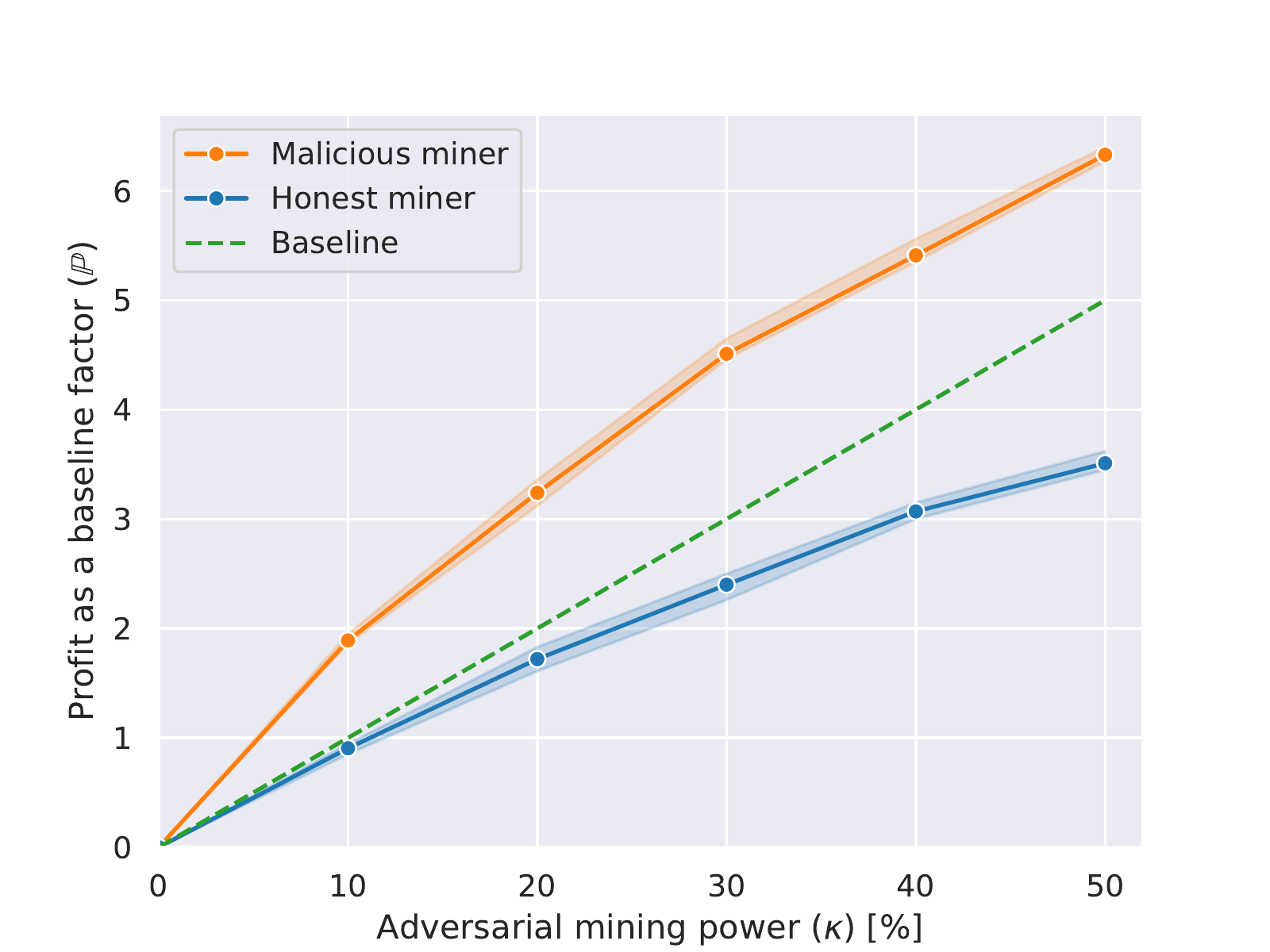}
	\caption{Profit of a single honest and malicious miner network power based on varying adversarial mining power \gls{adversarial-mining-power}, while \(\gls{block-creation-rate}=20\).}\label{fig:exp1-one-to-one}
\end{figure}

In all simulation runs, malicious miners earned profit disproportional to their mining power. 
Moreover, our experiments showed that the profit advantage of malicious miners decreases as their number increases. 
Intuitively, this happens because malicious miners increase transaction collision. In detail, since participants in PHANTOM are only rewarded for transactions that were first to be included in a new block, the profit for the second and later miner is lost if a duplicate transaction is included.

This observation might be seen as beneficial for the protocol as it disincentivizes multiple miners to use the rational transaction selection strategy.
On the other hand, we will show later that malicious miners are instead incentivized to create a malicious mining pool (see \autoref{sec:exp-3}).

Moreover, to supplement our observations, we performed another related experiment, in which we used the topology of one malicious and one honest miner, and we experimented with different adversarial mining power settings.
The results are depicted in \autoref{fig:exp1-one-to-one}, and they demonstrate that under such a setting, the malicious miner's relative earned profit (as compared to honest miner) grows proportionally to \gls{adversarial-mining-power}. 

\subsection{Experiment II}\label{sec:sec:exp-2}
\paragraph{Goal}
The goal of this experiment is to investigate the transaction throughput if the blockchain experiences malicious miners who select transactions based on a rational selection strategy.

\begin{figure}[t]
    \centering
    \includegraphics[width=\linewidth]{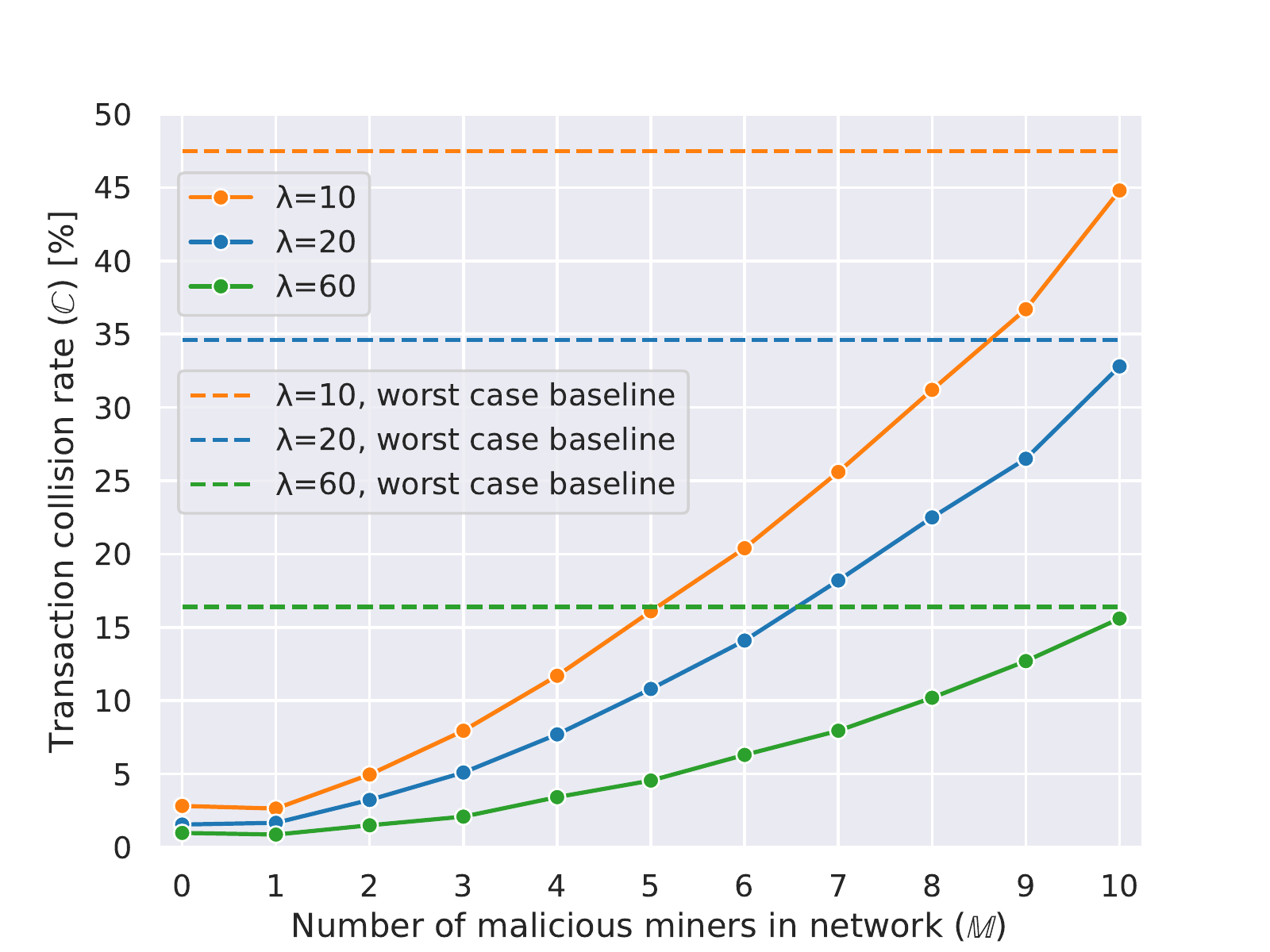}
    \caption{Transaction collision rate depending on the number of malicious miners (each with \(10\%\) of mining power) and \gls{block-creation-rate}. 
    The worst case scenario shows the transaction collision rate when all transactions in parallel blocks are duplicates.}\label{fig:transaction-collision-rate-proportional-to-amount-of-malicious-miners} 
\end{figure}

\begin{figure}[t]
    \centering
    \includegraphics[width=\linewidth]{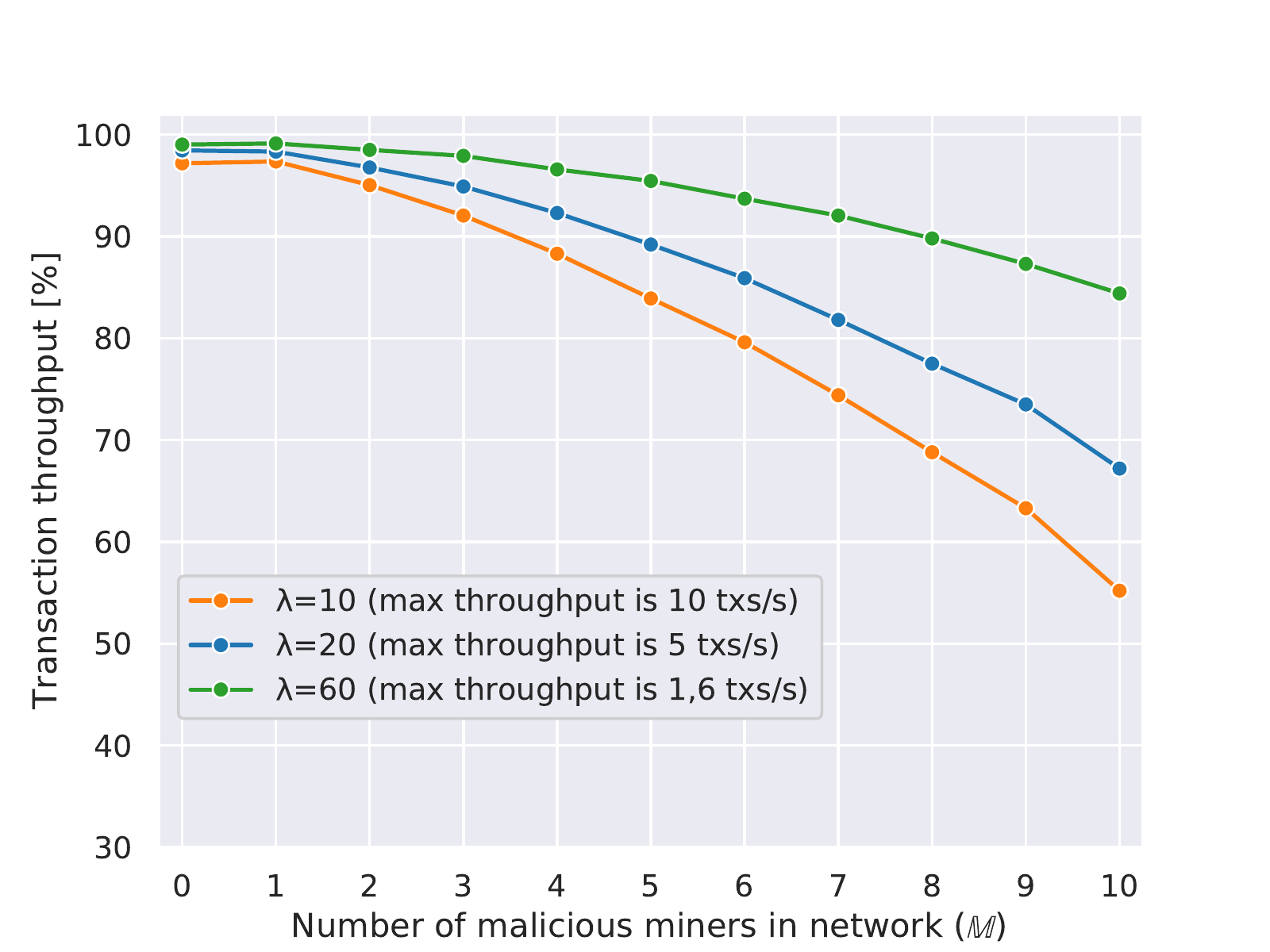}
    \caption{Throughput of the network depending on the number of malicious miners (each with 10\% of mining power) expressed as the percentage of all mined transactions that were not duplicate during a single simulation run.}\label{fig:scalability-disproportional-to-amount-of-malicious-miner} 
\end{figure}

\paragraph{Methodology and Results}
Alike in the previous experiment, the block creation time was kept constant (\gls{block-creation-rate}=20). 
The number of malicious miners (with 10\% of mining power) in the network was incrementally increased, and for each setting, the transaction collision rate and transaction throughput were computed. The results are depicted in \autoref{fig:transaction-collision-rate-proportional-to-amount-of-malicious-miners} and \autoref{fig:scalability-disproportional-to-amount-of-malicious-miner}, which are complementary. 
The figures confirm that the increase in the number of malicious miners also increases the transaction collision rate, which consequently decreases the overall throughput of the network.

\subsection{Experiment  III}\label{sec:exp-3}
\paragraph{Goal}
The goal of this experiment is to investigate the value of transaction collision rate under different \gls{block-creation-rate} and malicious miners present in the network.

\begin{figure}[t]
    \centering
    \includegraphics[width=\linewidth]{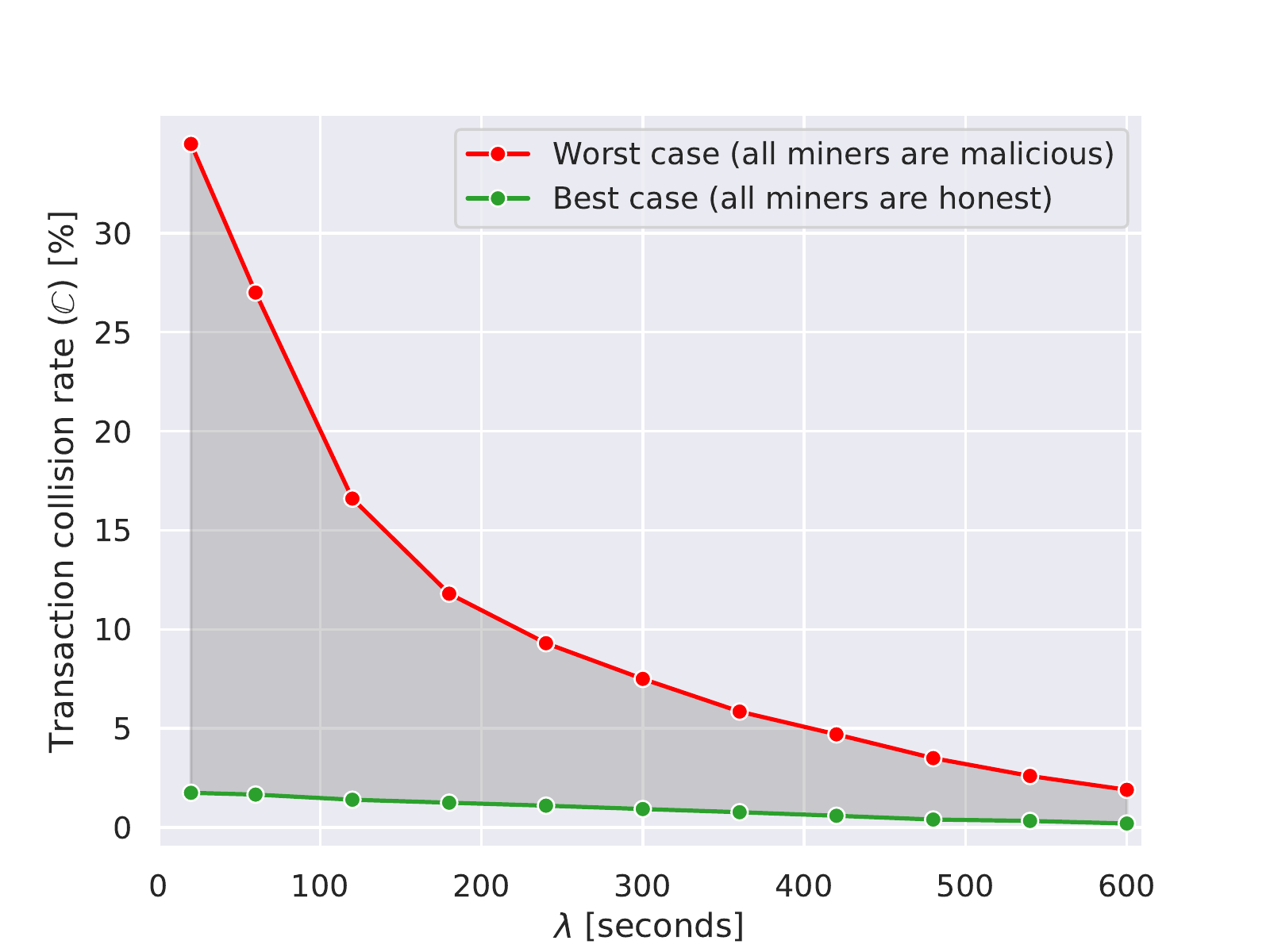}
    \caption{Transaction collision rate depending on \gls{block-creation-rate}, while all miners are either malicious or honest.}\label{fig:scalability-disproportional-to-block-creation-rate} 
\end{figure}

\begin{figure}[t]
	\centering
	\includegraphics[width=\linewidth]{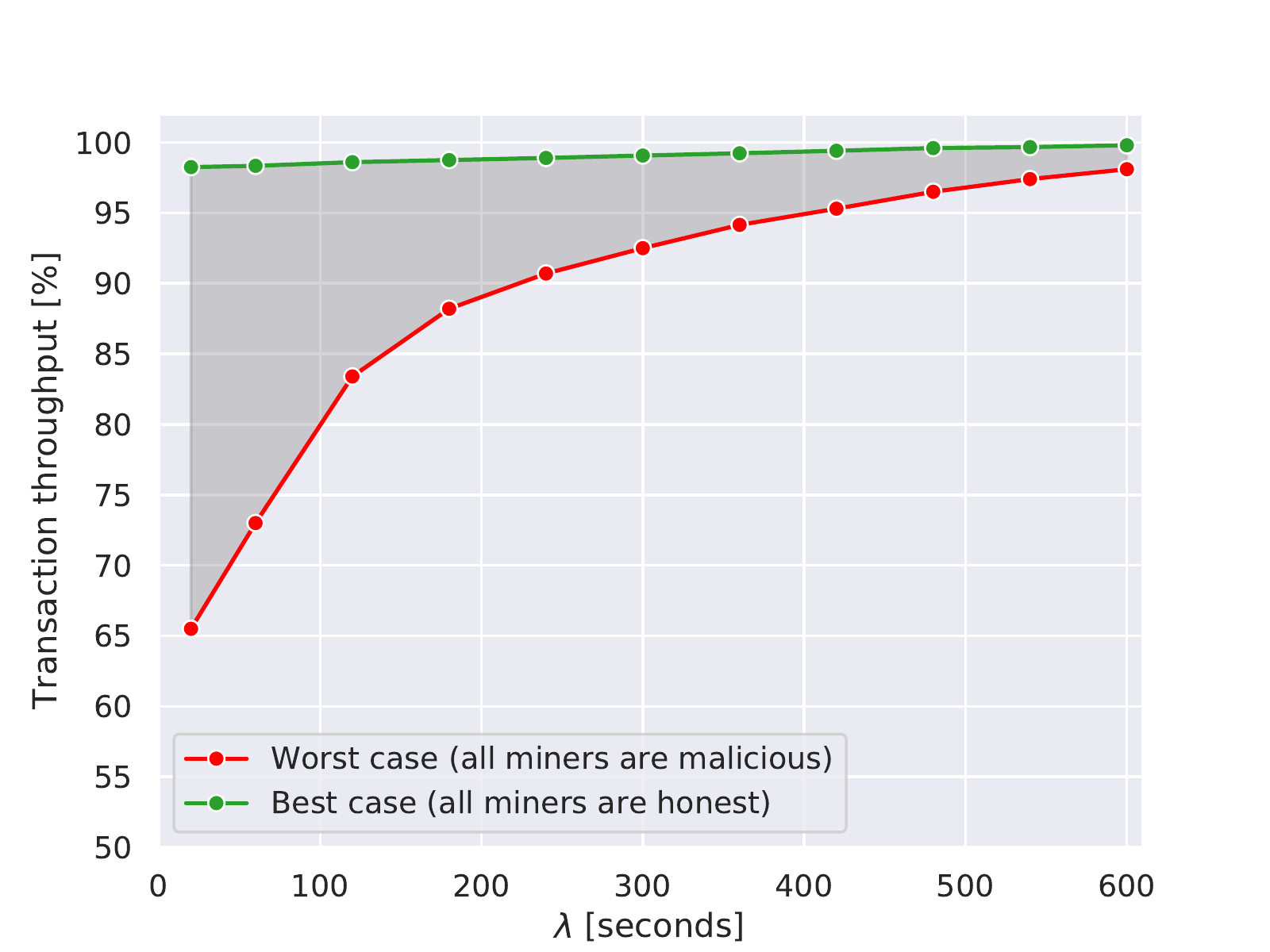}
	\caption{Transaction collision rate depending on \gls{block-creation-rate}, while all miners are either malicious or honest.}\label{fig:throughput-exp3}
\end{figure}

\paragraph{Methodology and Results}
We focused on two extreme scenarios, namely the scenario where all miners followed the random selection strategy (i.e., all miners were honest) and the scenario where all miners followed the rational selection strategy (i.e., all miners were malicious). 
The block creation time \gls{block-creation-rate} was changed within a specific interval. 
For each \gls{block-creation-rate}, we expressed the transaction collision rate as the percentage of all transactions mined during a single simulation run. The results are displayed in \autoref{fig:scalability-disproportional-to-block-creation-rate} and \autoref{fig:throughput-exp3}, which are complementary. 

The results confirm our expectations. First, when the random transaction selection strategy is followed by all nodes in the network, the transaction collision rate is negligible. Second, when all miners follow the rational transaction selection strategy, the transaction collision rate increases significantly with decreasing  \gls{block-creation-rate} since more blocks and transactions are produced at the same time. 
\section{Discussion and Future Work}\label{sec:discussion-and-future-work}

\paragraph{Centralization}
In the scope of Experiment III, we proved that the relative profit of malicious miners decreases as their number increases. Therefore, malicious miners are incentivized to form a single mining pool, 
maximizing their relative profit.
As a negative consequence, the decentralization of the blockchain network might be impacted. 

\paragraph{Throughput}
In our simulations, we adjusted the parameters not to focus on the maximum throughput of the simulated protocol but to investigate the potential issues that we discovered.
However, we argue that this had no impact on the principles of the simulated protocol, and similar results can be achieved even with higher throughputs.

\paragraph{Kaspa: The Implementation of GHOSTDAG}
Kaspa~\cite{Kaspa} is a real-world implementation of GHOSTDAG, which does not follow the originally proposed incentive scheme of random transaction selection, but selects transactions using a variant of rational strategy (i.e., taking into account the fees of transactions).
However, this might lead to the throughput problems identified in this work, and we conjecture that it is inefficient against rational miners as described in the work.

\paragraph{Future Work}
In our future work, we plan to study the influence of malicious miners on a network of larger size and different topology while considering various (higher) numbers of transactions in a block. 
Further, we plan to study the impact of malicious miners not only when they deviate from the protocol by changing the transaction selection strategy but also when they produce stale blocks as a result of a malicious action, which in turn requires the use of a particular discount function \gls{discount-function}.
In addition, we plan to propose and investigate the methods enabling \gls{dag}-oriented solutions such as PHANTOM and GHOSTDAG be resilient to demonstrated attacks, thus meeting the goal of increased throughput.

\section{Related Work}\label{sec:related}

\paragraph{\gls{dag}-Based Consensus Protocols}
The benefits of blockchain protocols come with certain trade-offs when balancing decentralization, scalability, and security. We have already mentioned the bottleneck in Nakamoto's consensus protocol. Thus, alternative approaches are emerging. We focus on the consensus protocols based on \gls{dag} that have been proposed so far.
Wang et al.~\cite{Wang2020SoKDI} performed a detailed systematic overview of such designs. They described six categories containing more than thirty \gls{dag}-based blockchain systems classified based on their characteristics and principles. They extend the commonly used classification based on the type of ledgers~\cite{Tang2020}.
GHOST~\cite{sompolinsky2015secure}, Inclusive Blockchain~\cite{lewenberg2015inclusive}, Conflux~\cite{li2018scaling}, Haootia~\cite{Tang2020}, and Byteball~\cite{Churyumov2016} represent \gls{dag} with the main chain. 
Ledgers with parallel chains are represented by Hashgraph~\cite{hashgraph2016} and Nano~\cite{Lemahieu2018NanoA}. 

Nevertheless, our focus is purely on DAGs with a main chain, such as SPECTRE~\cite{sompolinsky2016spectre}, PHANTOM and GHOSTDAG~\cite{sompolinsky2020phantom}. For details about these protocols, we refer the reader to \autoref{sec:background}.

\paragraph{Performance Analysis of DAGs}
While many manuscripts deal with the security and performance analysis of mentioned protocols, they consider neither mining strategy nor features of various incentive schemes. 
Park et al.~\cite{8756973} address the performance of DAG-based blockchains and relate it to the optimization of profit. They show that the average number of parents \(n\) can influence the transaction processing time and also transaction speed. As a result, they propose a competitive-based transaction process system using a dynamic fee policy.

Birmpas et al.~\cite{Birmpas2020FairnessAE} propose a new general framework that captures ledger growth for a large class of DAG-based implementations to demonstrate structural limits of DAG-based protocols. Even with honest miner behaviour, fairness and efficiency of the ledger can be affected by to different connectivity levels.

One of the key technical problems of DAG-based protocols is the identification of honest blocks. Wang proposes a MaxCord~\cite{MaxCord}, a framework using a different approach for honest block identification problem using graph theory. Based on the definition of the disparity measurement between blocks, they convert the problem into a maximum k-independent set problem.
Cao et al.~\cite{CAO2020480} compared the performance of three consensus mechanisms: Bitcoin (PoW), Nxt (PoS), and Tangle (\gls{dag}-PoW) in terms of parameters such as average time to generate a new block or confirmation delay and failure probability, showing how these mechanisms can affect the state of network resources or network load condition. 
\section{Conclusion}\label{sec:conclusion}
In this work, we have provided an overview of \gls{dag}-oriented protocols that promise to increase the transaction throughput on the blockchain network. Emphasis was given to the PHANTOM protocol and its optimization GHOSTDAG. 
In our experiments, we analyzed the impact of malicious miners who deviate from the PHANTOM protocol by selecting transactions for new candidate blocks based on the highest fee rather than randomly (as proposed in the original design).

Within three dedicated experiments, we proved that malicious actors have a significant advantage over honest actors in terms of profit maximization, even in the situation where protocol security has been sacrificed for the sake of profit maximization of honest miners (i.e., where a significant advantage was given to honest miners).
This implies that the influence of malicious miners would only be magnified in a real-world scenario. Moreover, we show that malicious actors have a detrimental impact on transaction throughput and have a significant incentive to form a mining pool, sacrificing the decentralization of the solution. 

\bibliographystyle{IEEEtran}
\bibliography{ref}

\printglossaries{}

\end{document}